\documentclass{ws-procs975x65}

\newcommand{\vev}[1]{\langle #1\rangle}

\begin{document}

\title{Recent variations of fundamental parameters and their 
implications for gravitation}

\author{Thomas Dent\footnote{E-mail: thomas.dent@astro.cardiff.ac.uk}\footnote{The collaboration of S.~Stern and C.~Wetterich (University of Heidelberg) in parts of this work is acknowledged.}}

\address{Physics \& Astronomy, Cardiff University,\\
Cardiff CF24 3AA, U.K.\\
}



\begin{abstract}
We compare the sensitivity of a recent bound on time variation of the fine structure constant from optical clocks with bounds on time varying fundamental constants from atomic clocks sensitive to the electron-to-proton mass ratio, from radioactive decay rates in meteorites, and from the Oklo natural reactor. Tests of the Weak Equivalence Principle also lead to comparable bounds on present time variations of constants, as well as putting the strongest limits on variations tracking the gravitational potential. For recent time variations, the ``winner in sensitivity'' depends on possible relations between the variations of different couplings in the standard model of particle physics. WEP tests are currently the most sensitive within scenarios with unification of gauge interactions. A detection of time variation in atomic clocks would favour dynamical dark energy and put strong constraints on the dynamics of a cosmological scalar field. 
\end{abstract}


\bodymatter

\paragraph{Introduction: Temporal and local variation}
Both the Standard Model of particle physics and the theory of General Relativity are constructed on the assumption of Local Position Invariance (LPI): that locally measurable properties of matter do not vary over space and time. LPI forms part of Einstein's equivalence principle. While most tests of LPI concerning particle coupling strengths and masses have yielded results consistent with zero variation \cite{Uzan:2002vq}, there is a significant indication of cosmological variation in the fine structure constant $\alpha$ deduced from astrophysical absorption spectra \cite{Murphy:2003mi,Murphy:2003hw}, driving recent interest in the possibility of nonzero variations. The latest result from spectroscopy of a large sample is $\Delta \alpha/\alpha = (-0.57 \pm 0.11) \times 10^{-5}$ over a range $0.2<z_{abs}<4.2$.

The proton-electron mass ratio $\mu\equiv m_p/m_e$ has also been probed: earlier analyses of molecular H$_2$ spectra indicated a moderately significant variation \cite{Reinhold:2006zn,Ivanchik:2005ws} at high redshift ($z\sim 2$-$3$). Recently, a strong bound on variation at lower redshift has been derived from the NH$_3$ inversion spectrum \cite{Flambaum:2007fa,Murphy:2008yy} with a 95\% confidence level limit of $\Delta\mu/\mu < 1.8\times 10^{-6}$ quoted at $z=0.685$. 

Time variation of couplings and masses in the recent Universe is stringently limited by atomic clock measurements. Evolution of $\alpha$ is bounded by direct comparison of optical frequencies in an Al/Hg ion clock \cite{Rosenband}:
\begin{equation} \label{eq:alphabound}
 d \ln \alpha / dt = (-1.6 \pm 2.3) \times 10^{-17} y^{-1},
\end{equation}
while the limit on the proton-electron mass ratio from 
a recent evaluation \cite{Blatt:2008su} of atomic clock data is
\begin{equation} \label{eq:mubound}
  d \ln \mu / dt = (1.5 \pm 1.7) \times 10^{-15} y^{-1} .
\end{equation}
It is also of interest to probe possible variations at the present time within the Solar System: any theory with an underlying Lorentz invariance which allows time variation, should also allow spatial variation. 
Recently bounds have been set \cite{Blatt:2008su,Fortier:2007jf} on variations of fundamental parameters correlated with the gravitational potential $U$, by comparing atomic clock frequencies over several months while the Earth moves through the Sun's gravitational field \cite{Flambaum:2007ar,Shaw:2007ju}. For independently varying fundamental parameters $G_i$ we may define couplings $k_i$ via
\begin{equation} \label{dkdef}
	\Delta \ln G_i = k_i \Delta U
\end{equation}
for small changes in $U$; the annual variation due to Earth's orbital eccentricity is $\Delta U\simeq 3\times 10^{-10}$. 
The couplings $k$ of $\alpha$, $\mu$ and the light quark mass $m_q/\Lambda_c\equiv (m_u+m_d)/2\Lambda_c$, where $m_u$ and $m_d$ are the up and down quark current masses and $\Lambda_c$ is the invariant strong interaction scale of QCD, 
were found to be consistent with zero \cite{Blatt:2008su}, with 
uncertainties
\begin{equation}
	\{\sigma(k_\alpha),\sigma(k_\mu),\sigma(k_q)\}= \{3.1,17,27\}\times 10^{-6}.
\end{equation} 
Using the results of \cite{Rosenband} an improved limit can be set \cite{Barrow:2008se} giving $k_\alpha = (-5.4\pm 5.1)\times 10^{-8}$.

\paragraph{Weak Equivalence Principle violation and local variation}
However, any spatial gradient of couplings or particle masses leads to anomalous accelerations of test bodies\cite{PeeblesDicke}.
We have 
$	\vec{\nabla}\ln G_i = - k_i \vec{g}$,
where $\vec{g} = -\vec{\nabla}U$.
A freely-falling body of mass $M(G_i(\vec{x}))$ will experience additional acceleration $\vec{a}=-(\vec{\nabla} M)/M$ \cite{Brans:1961sx}, as if moving in a potential $V(\vec{x})= M(G_i(\vec{x}))$. For test bodies of different $\bf{x}$-dependence the acceleration cannot be absorbed into a redefinition of the metric, and the {\em differential}\/ acceleration $\vec{a}_b-\vec{a}_c$, violating the Weak Equivalence Principle (WEP) or universality of free fall, is a physical signal probing gradients of $G_i$.

Using Planck units where $G_{\rm N}$ is constant. the differential acceleration is then \cite{Damour:1994zq}
\begin{equation}
	\eta_{b-c}\equiv \frac{|\vec{a}_b-\vec{a}_{c}|}{|\vec{g}|}
	= \sum_i \frac{\partial \ln (M_b/M_c)}{\partial \ln G_i} k_i 
	\equiv \sum_i \lambda^{b-c}_i k_i.
\end{equation}
which defines the sensitivity coefficients $\lambda^{b-c}_i$. Each E{\" o}tv{\" o}s-type experimental limit on $\eta$\cite{Schlamminger:2007ht,Baessler:1999iv,Su:1994gu,Braginsky72} bounds some linear combination of couplings $k_i$.

To calculate $\lambda^{b-c}_i$ we first consider dimensionless ``nuclear parameters'' which characterize physics at low energy: 
the fine structure constant $\alpha$; 
the electron mass $m_e/m_N$; 
the nucleon mass difference $\delta_N/m_N$;  
and the nuclear surface tension $a_S/m_N$, 
where $\delta_N\equiv m_n-m_p$ and $m_N\equiv (m_n+m_p)/2$. E{\" o}tv{\" o}s experiments cannot distinguish between couplings to $\delta_N/m_N$ and $m_e/m_N$, since both lead to forces proportional to the proton fraction $f_p \equiv Z/A$: hence we define $Q_n\equiv \delta_N-m_e$. Null bounds were then found \cite{Dent:2008gu} on three couplings to $U$, with uncertainties
\begin{equation} 
	\{\sigma(k_{Qn}),\sigma(k_\alpha),\sigma(k_{aS})\} = 
	\{38, 2.3, 1.0\}\times 10^{-9}.
\end{equation}
Estimating how the ``nuclear parameters'' $(Q_n/m_N,\alpha,a_S/m_N)$ depend on fundamental Standard Model (SM) parameters $\alpha$, 
$m_e/\Lambda_c$, 
average light quark mass $m_q/\Lambda_c$, 
and up-down mass difference $\delta_q/\Lambda_c \equiv (m_d-m_u)/\Lambda_c$, 
where $\Lambda_c$ is the QCD strong interaction scale, we find \cite{Dent:2008gu} the bounds 
\begin{equation}
	\{\sigma(k'_{\delta f}),\sigma(k'_{\alpha}),\sigma(k'_q)\} = 
	\{14,1.7,0.9\}\times 10^{-9}. 
\end{equation}
Here $k'$ denote the couplings to $U$ in the basis of independently varying {\em fundamental}\/ parameters, and $k'_{\delta f} \equiv k'_{\delta q}-0.25k'_{e}$. Hence WEP tests provide 
the most stringent bounds on variations of fundamental couplings correlated with the gravitational potential \cite{Nordtvedt:2002qe}.

\paragraph{Cosmological and recent time dependence} Returning to possible time variations, one can study the dependence over cosmologically long periods \cite{Dent:2008vd} $z\gtrsim1$, which however implies a strongly model-dependent extrapolation in comparing to present-day variations. For more recent Òhistorical boundsÓ including nuclear decay rates in meteorites \cite{Olive:2003sq,Dent:2008gx} and isotopic abundances in the Oklo natural reactor \cite{Damour:1996zw, Fujii:1998kn, Petrov:2005pu, Gould:2007au}, the time span to the present is short on cosmological scales and a linear time interpolation is meaningful in some scenarios of evolution. On this basis we can compare the present-day limits \eqref{eq:alphabound} and \eqref{eq:mubound} with `recent' bounds and with the sensitivity of bounds on time variations from WEP tests \cite{Dent:2008ev}. 

Such comparisons involve assumptions about relations between the variations of different SM parameters $G_k = \{G_{\rm N}, \alpha, \langle\phi\rangle, m_e, \delta_q, m_q \}$, where $\langle\phi\rangle$ is the Higgs v.e.v.\footnote{We take units where $\Lambda_c$ is constant.}. The comparison with WEP bounds will also use cosmological constraints on the time evolution of a scalar field. 

We suppose that fractional variations of $G_k$ are proportional to one nontrivial variation, with fixed constants of proportionality, over a recent cosmological period $z\lesssim 0.5$, due to the existence of a Grand Unified field theory (GUT) \cite{Calmet:2001nu,Calmet:2002ja,Langacker:2001td}. 
We write 
\begin{equation} 
 \Delta \ln G_k = d_k \Delta \ln \alpha_X
\end{equation}
for some constants $d_k$ which depend on the specific GUT model. Assuming constant Yukawa couplings, the electron and quark masses simply vary proportional to the Higgs v.e.v.\ leaving each {\em unified scenario}\/ to be defined by the variations of $M_X/M_{\rm P}$ ($M_X$ being the unification mass scale), $\alpha_X$, $\langle\phi\rangle/M_X$, and $m_S/M_X$, where $m_S$ is the mass scale of supersymmetric partners of SM particles, if they exist. We defined a number of scenarios with different proportionality constants $d_{M,X,H,S}$ motivated by particle physics models \cite{Dent:2008gx} both with and without supersymmetry. 

In Table~\ref{tab:PresentVariationErrors} we compare the precision of bounds on fractional variations of the fundamental parameter in each scenario ($\alpha$, $\alpha_X$ or $\vev{\phi}/M_X$). \footnote{In addition to unified models we have considered the case where only $\alpha$ varies.} The column ``Clocks'' results from atomic clock bounds on $\mu$, the recent Al/Hg limit \cite{Rosenband} on $\alpha$ variation being treated separately. The Oklo bound on $\alpha$ variation is rescaled to account for systematic uncertainties \cite{Dent:2008ev}.
\begin{table}
\tbl{Competing bounds on present time variations. For each scenario we give the $1\sigma$ uncertainties of null bounds on $d(\ln X)/dt$.}
{\begin{tabular}{lc|ccccc}
\toprule
 & & \multicolumn{5}{c}{ Error on $d \ln X/dt$ ($10^{-15}y^{-1}$)} \\
\hline
Scenario & $X$     & Al/Hg & Clocks ($\mu$) & Oklo & ${^{187}}$Re & WEP \\
\hline
$\alpha$ only & $\alpha$    &  0.023 & -      & 0.033 & 0.32  & 6.2    \\
2        & $\alpha_X$       & 0.027  & 0.074  & 0.12  & 0.015 & 0.007 \\
2S       & $\alpha_X$       & 0.044  & 0.12   & 0.19  & 0.026 & 0.012 \\
3        & $\vev{\phi}/M_X$ & 12.4   & 2.6    & 54    & 0.53  & 0.33   \\
4        & $\vev{\phi}/M_X$ & 1.78   & 6.2    & 7.7   & 1.2   & 0.35   \\
5, $\tilde{\gamma} = 42$ & $\alpha_X$ & 0.024  & 0.42  & 0.11  & 0.069 & 0.013  \\
6, $\tilde{\gamma} = 70$ & $\alpha_X$ & 0.016  & 0.30  & 0.070 & 0.049 & 0.008 \\
6, $\tilde{\gamma} = 25$ & $\alpha_X$ & 0.027  & 0.25  & 0.12  & 0.056 & 0.011 \\
\botrule
\end{tabular}}
\label{tab:PresentVariationErrors}
\vspace*{0.2cm}
\end{table}
Considering Al/Hg, other clocks, Oklo and the ${^{187}}$Re decay from meteorites, if only $\alpha$ varies, atomic clock experiments are already the most sensitive, surpassing the previous best bound from the Oklo reactor. If only the unified coupling $\alpha_X$ varies, the meteorite bound is still somewhat stronger than atomic clocks. The same holds if only the ratio $\vev{\phi} / M_X$ varies. For combined variations of $\alpha_X$ and $\vev{\phi}/M_X$ in scenarios 5 and 6, the ``sensitivity winner'' is again the laboratory bound. Thus, laboratory experiments have now reached the sensitivity of the ``historical'' bounds; a modest further increase in sensitivity, especially for variations of $\mu$, will make them the best probes in all scenarios.
\vspace*{-6.5mm}
\paragraph{Quintessence and violation of the WEP} 
In quantum field theory, any time variation of couplings must be associated to the time evolution of a field, which may be fundamental or a composite operator. The simplest hypothesis is a scalar field whose time-varying expectation value preserves rotation and translation symmetry locally, as well as all gauge symmetries of the standard model. We may identify this scalar field with the ``cosmon'' field $\varphi$ of dynamical dark energy or quintessence models. 

Table \ref{tab:PresentVariationErrors} shows a further competitor for the ``sensitivity race'': bounds on WEP violation. In order to cause nontrivial variation the cosmon couples non-universally to SM fields; thus it mediates a ``fifth force''. 
Two test bodies with 
different composition will experience different accelerations towards a common source, due to their generally different ``cosmon charge'' per mass. 

The cosmon couplings to atoms and photons necessarily result in {\em spatial}\/ variation in the vicinity of concentrations of matter, which may cause seasonal variations as discussed above.
However spatial variations over cosmological scales generally do not influence the slow evolution over time \cite{Shaw:2005vf} that we focus on.

The cosmon couplings to atoms and photons determine both the outcome of tests of the WEP, and the time variation of ``constants'' at present and in recent cosmological epochs. 
To bound time-varying parameters from WEP tests one needs information on the rate of change of the expectation value of the scalar field.
The latter can be expressed in terms of cosmological observables, namely the fraction in dark energy contributed by the cosmon, $\Omega_h$, and its equation of state, $w_h$, as $\dot{\varphi}^2 / 2 = \Omega_h (1 + w_h) \rho_c$. Here $\rho_c = 3 H^2 M_P^2$ is the critical energy density of the Universe. 

The differential acceleration $\eta$ 
can now be related to the present time variation of couplings and to cosmological parameters. Taking the fine structure constant as the varying parameter, we find  \cite{Wetterich:2002ic, 
Dent:2008vd} at the present epoch
\begin{equation}
\label{eq:alphacosmoeta}
 \left| \frac{\dot{\alpha}/\alpha}{10^{-15}{\rm y}^{-1}} \right| = \left|\frac{\Omega_h (1 + w_h)}{F} \right|^{1/2} \left| \frac{\eta}{3.8 \times 10^{-12}} \right|^{1/2} .
\end{equation}
The ``unification factor'' $F$ encodes the dependence on the precise unification scenario 
via the coupling strengths of the cosmon, and on the composition of the test bodies. The values of $F$ were computed  
for the 
best current test of WEP \cite{Schlamminger:2007ht} yielding $\eta = (0.3\pm 1.8)\times 10^{-13}$ for Be-Ti test masses, and are shown in 
\begin{table}
\tbl{Values of $F$ for the WEP experiment of  \cite{Schlamminger:2007ht} in different unified scenarios.}
{\begin{tabular}{c||c|c|c|c|c|c|c}
\hline
Scenario & $\alpha$ only & 2 & 3 & 4 & 5, $\tilde{\gamma}\! =\! 42$ & 6, $\tilde{\gamma}\! =\! 70$ & 6, $\tilde{\gamma}\! =\! 25$ \\ 
\hline 
$F$ (Be-Ti)& $-9.3\, 10^{-5}$ & 95  & -9000 & -165  & -25  & -26  & 41 \\
\hline
\end{tabular}}
\label{tab:F}
\end{table}
Table~\ref{tab:F}. 
%
The relation \eqref{eq:alphacosmoeta} then allows for a direct comparison between the sensitivity of measurements of $\eta$ with measurements of $\dot{\alpha}/\alpha$ from laboratory experiments, or bounds from recent cosmological history
provided we can use cosmological information for $\Omega_h (1 + w_h)$. If the time varying field is responsible for the dark energy in the Universe, then cosmological observations imply $\Omega_h \simeq 0.73$ and $w_h\leq -0.9$.

The resulting bounds 
are displayed in the last column of Tab.~\ref{tab:PresentVariationErrors}. For all unification scenarios these bounds are the most severe;
At present, clock experiments win the ``sensitivity race'' only if $\alpha$ is the only time-varying coupling. 
WEP bounds become even more restrictive if the scalar field plays no significant role in cosmology. In this case $\Omega_h$ is small, say $\Omega_h < 0.01$. 
The relevant limit is an observational bound on the scalar kinetic energy
independent of the precise scalar model and of its overall role within cosmology. 

The converse argument is equally strong: if a time variation of the fine structure constant is observed close to the present limit, most unification scenarios can be ruled out. Only scenarios with a value $|F| \lesssim 10$ would remain compatible with the WEP bound from Be/Ti test masses.
Further precise tests with materials of different composition will make it even more difficult to ``hide'' the cosmon coupling to photons (needed for $\dot{\alpha}/\alpha \neq 0$) from detection in WEP tests. 

A 
nonzero variation near the present experimental bounds would put important constraints on cosmology: it would require some field to be evolving in the present cosmological epoch, 
ruling out the minimal $\Lambda$CDM model and favouring dynamical dark energy models. 
Furthermore, a combination of time-varying couplings with WEP bounds on $\eta$ would provide a {\em lower bound} on the kinetic energy of dynamical models.
For a scalar field we find the bound 
\begin{equation}
\label{eq:BoundOnOmegaw}
 \Omega_h (1+w_h) \gtrsim 3.8 \times 10^{18} F (\partial_t \ln \alpha)^2 
 \eta_{\rm max}^{-1} \, .
\end{equation}
Here $\partial_t \ln \alpha$ is given in units of y$^{-1}$, and $\eta_{\rm max}$
is the current experimental limit on WEP violation.
Thus, if $|\partial_t \ln \alpha |$ is nonzero and not too small, $w_h$ {\em cannot}\/ be arbitrarily close to $-1$ (a cosmological constant); nor can the contribution of the scalar to the dark energy density be insignificant.

The race for the best bounds on the present time variation of fundamental couplings is open, independently of other possible interesting observations of varying couplings in early cosmological epochs. The precision of WEP-bounds on $\eta$ and atomic clock bounds on $\dot{\alpha}/\alpha$ and $\dot{\mu}/\mu$ is expected to increase. 
The largest restrictive power, or the largest discovery potential for nontrivial cosmological dynamics, arises from a combined increase in sensitivity of both WEP and clock experiments. 

\bibliographystyle{ws-procs975x65}


\begin{thebibliography}{10}

\bibitem{Uzan:2002vq}
J.-P. Uzan, {\em Rev. Mod. Phys.} {\bf 75}, p. 403 (2003).

\bibitem{Murphy:2003mi}
M.~T. Murphy {\em et~al.}, {\em Lect. Notes Phys.} {\bf 648}, 131 (2004).

\bibitem{Murphy:2003hw}
M.~T. Murphy, J.~K. Webb and V.~V. Flambaum, {\em Mon. Not. Roy. Astron. Soc.}
  {\bf 345}, p. 609 (2003).

\bibitem{Reinhold:2006zn}
E.~Reinhold {\em et~al.}, {\em Phys. Rev. Lett.} {\bf 96}, p. 151101 (2006).

\bibitem{Ivanchik:2005ws}
A.~Ivanchik {\em et~al.}, {\em Astron. Astrophys.} {\bf 440}, 45 (2005).

\bibitem{Flambaum:2007fa}
V.~V. Flambaum and M.~G. Kozlov, {\em Phys. Rev. Lett.} {\bf 98}, p. 240801
  (2007).

\bibitem{Murphy:2008yy}
M.~T. Murphy, V.~V. Flambaum, S.~Muller and C.~Henkel, {\em Science} {\bf 320},
  1611 (2008).

\bibitem{Rosenband}
T.~Rosenband {\em et~al.}, {\em Science} {\bf 319}, 1808 (2008).

\bibitem{Blatt:2008su}
S.~Blatt {\em et~al.}, {\em Phys. Rev. Lett.} {\bf 100}, p. 140801 (2008).

\bibitem{Fortier:2007jf}
T.~M. Fortier {\em et~al.}, {\em Phys. Rev. Lett.} {\bf 98}, p. 070801 (2007).

\bibitem{Flambaum:2007ar}
V.~V. Flambaum and E.~V. Shuryak, {\em AIP Conf. Proc.} {\bf 995}, 1 (2008).

\bibitem{Shaw:2007ju}
D.~J. Shaw (2007), eprint gr-qc/0702090.

\bibitem{Barrow:2008se}
J.~D. Barrow and D.~J. Shaw, {\em Phys. Rev.} {\bf D78}, p. 067304 (2008).

\bibitem{PeeblesDicke}
P.~Peebles and R.~Dicke, {\em Phys. Rev.} {\bf 127} (1962), see also T.~Damour,
  gr-qc/9606079.

\bibitem{Brans:1961sx}
C.~Brans and R.~H. Dicke, {\em Phys. Rev.} {\bf 124}, 925 (1961).

\bibitem{Damour:1994zq}
T.~Damour and A.~M. Polyakov, {\em Nucl. Phys.} {\bf B423}, 532 (1994).

\bibitem{Schlamminger:2007ht}
S.~Schlamminger, K.-Y. Choi, T.~Wagner, J.~Gundlach and E.~Adelberger, {\em
  Phys. Rev. Lett.} {\bf 100}, p. 041101 (2008).

\bibitem{Baessler:1999iv}
S.~Baessler {\em et~al.}, {\em Phys. Rev. Lett.} {\bf 83}, p. 003585 (1999).

\bibitem{Su:1994gu}
Y.~Su {\em et~al.}, {\em Phys. Rev.} {\bf D50}, 3614 (1994).

\bibitem{Braginsky72}
V.~G. Braginsky and V.~I. Panov, {\em Sov. Phys. JETP} {\bf 34}, p. 463 (1972).

\bibitem{Dent:2008gu}
T.~Dent, {\em Phys. Rev. Lett.} {\bf 101}, p. 041102 (2008).

\bibitem{Nordtvedt:2002qe}
K.~Nordtvedt, {\em Int. J. Mod. Phys.} {\bf A17}, 2711 (2002).

\bibitem{Dent:2008vd}
T.~Dent, S.~Stern and C.~Wetterich, {\em JCAP} {\bf 0901}, p. 038 (2009).

\bibitem{Olive:2003sq}
K.~A. Olive {\em et~al.}, {\em Phys. Rev.} {\bf D69}, p. 027701 (2004).

\bibitem{Dent:2008gx}
T.~Dent, S.~Stern and C.~Wetterich, {\em Phys. Rev.} {\bf D78}, p. 103518
  (2008).

\bibitem{Damour:1996zw}
T.~Damour and F.~Dyson, {\em Nucl. Phys.} {\bf B480}, 37 (1996).

\bibitem{Fujii:1998kn}
Y.~Fujii {\em et~al.}, {\em Nucl. Phys.} {\bf B573}, 377 (2000).

\bibitem{Petrov:2005pu}
Y.~V. Petrov, A.~I. Nazarov, M.~S. Onegin, V.~Y. Petrov and E.~G. Sakhnovsky,
  {\em Phys. Rev.} {\bf C74}, p. 064610 (2006).

\bibitem{Gould:2007au}
C.~R. Gould, E.~I. Sharapov and S.~K. Lamoreaux, {\em Phys. Rev.} {\bf C74}, p.
  024607 (2006).

\bibitem{Dent:2008ev}
T.~Dent, S.~Stern and C.~Wetterich, {\em Phys. Rev.} {\bf D79}, p. 083533
  (2009).

\bibitem{Calmet:2001nu}
X.~Calmet and H.~Fritzsch, {\em Eur. Phys. J.} {\bf C24}, 639 (2002).

\bibitem{Calmet:2002ja}
X.~Calmet and H.~Fritzsch, {\em Phys. Lett.} {\bf B540}, 173 (2002).

\bibitem{Langacker:2001td}
P.~Langacker, G.~Segre and M.~J. Strassler, {\em Phys. Lett.} {\bf B528}, 121
  (2002).

\bibitem{Shaw:2005vf}
D.~J. Shaw and J.~D. Barrow, {\em Phys. Lett.} {\bf B639}, 596 (2006).

\bibitem{Wetterich:2002ic}
C.~Wetterich, {\em JCAP} {\bf 0310}, p. 002 (2003).

\end{thebibliography}

\end{document}